\documentclass[floatfix,reprint,nofootinbib,amsmath,amssymb,epsfig,aip,floats,letterpaper,groupedaffiliation,onecolumn]{revtex4-1}

\usepackage{amsmath}
\usepackage{amssymb}
\usepackage{graphicx}
\usepackage{epstopdf}
\usepackage{hyperref}
\usepackage{dcolumn}
\usepackage{bm}
\usepackage{listings}
\usepackage{xcolor}
\usepackage{inconsolata}

\colorlet{punct}{red!60!black}
\definecolor{background}{HTML}{EEEEEE}
\definecolor{delim}{RGB}{20,105,176}
\colorlet{numb}{magenta!60!black}

\lstdefinelanguage{json}{
    basicstyle=\fontsize{9}{12}\ttfamily,
    showstringspaces=false,
    breaklines=true,
    literate=
     *{0}{{{\color{numb}0}}}{1}
      {1}{{{\color{numb}1}}}{1}
      {2}{{{\color{numb}2}}}{1}
      {3}{{{\color{numb}3}}}{1}
      {4}{{{\color{numb}4}}}{1}
      {5}{{{\color{numb}5}}}{1}
      {6}{{{\color{numb}6}}}{1}
      {7}{{{\color{numb}7}}}{1}
      {8}{{{\color{numb}8}}}{1}
      {9}{{{\color{numb}9}}}{1}
      {:}{{{\color{punct}{:}}}}{1}
      {,}{{{\color{punct}{,}}}}{1}
      {\{}{{{\color{delim}{\{}}}}{1}
      {\}}{{{\color{delim}{\}}}}}{1}
      {[}{{{\color{delim}{[}}}}{1}
      {]}{{{\color{delim}{]}}}}{1},
}

\newcommand{\beq}{\begin{equation}}
\newcommand{\eeq}{\end{equation}}

\begin{document}

\title{Sidecoin: a Snapshot Mechanism for Bootstrapping a Blockchain}

\author{Joseph Krug \& Jack Peterson\\www.sidecoin.net}

\begin{abstract}
Sidecoin is a mechanism that allows a snapshot to be taken of Bitcoin's blockchain. We compile a list of Bitcoin's unspent transaction outputs, then use these outputs and their corresponding balances to bootstrap a new blockchain. This allows the preservation of Bitcoin's economic state in the context of a new blockchain, which may provide new features and technical innovations.
\end{abstract}

\maketitle

\section{Snapshots and Their Uses}

A snapshot records all public addresses in the Bitcoin network with a substantial balance (over about 20 cents in USD) and sorts them from highest to lowest balance. This eliminates millions of addresses from places like Satoshi Dice that have small amounts of bitcoin in them, called `dust'. We then save these addresses, along with their balances, and an alternative representation of the address known as a hash160.\footnote{`hash160' is shorthand for the 160-bit hash created by SHA256 followed by RIPEMD160.}

Snapshots allow us to do a few things.  They can be used to examine addresses and balances to find high value holders, or to see the distribution of wealth among Bitcoiners: the economic state of the Bitcoin network. However, a much more fun use is to create altcoins -- or, as coins created using this method are called, \emph{spinoff} coins.\footnote{\texttt{http://bitcointalk.org/index.php?topic=563972.0}} To create a spinoff coin, we first take a snapshot of Bitcoin's blockchain, then create an altcoin with an initial distribution based on Bitcoin's. \emph{Sidecoin} is the reference implementation for bootstrapping a new blockchain using Bitcoin's present state.  This technology allows bitcoin owners to trustlessly claim a proportional amount of a new cryptocurrency.

Imagine a coin development team like Gridcoin's\footnote{\texttt{http://www.gridcoin.us}}.  Gridcoin uses a proof-of-work that, in principle, benefits society through computation, by directing it to to fold proteins for cancer research. Their team would probably be pleased if a tiny bit of the enormous mining power used to mine bitcoins could be redirected to their \emph{useful} proof-of-work system. It provides a principled, transparent distribution method with which `pre-mined' cryptocurrencies can disseminate their tokens.  One example is Ripple, which does not use mining at all. Taking a snapshot, and allowing Bitcoiners to claim coins on the Gridcoin or Ripple networks, removes the profit gained by releasing new cryptocurrencies which do not contribute new technology.  Coin developers can still profit by allowing mining after the initial claiming, but only by providing a true improvement to Bitcoin or other altcoins.

A snapshot-based distribution bootstraps an altcoin with a much larger user base than they otherwise would have had: any bitcoin owner is automatically `bought-in' to the altcoin, provided they decide to claim their coins. Bitcoin holders who think an altcoin does not provide sufficient technological advancements over Bitcoin can simply claim their new coins, then sell them.  Altcoin creators who believe that their coins hold value due to useful features are incentivized to buy more after the initial sharp selloff, because they will profit once the market realizes the value of their innovations.

\section{Taking a Snapshot}

There were several steps involved in creating our snapshot file.  First, we downloaded the Bitcoin blockchain using \texttt{bitcoind}, then we parsed all the balances in the Bitcoin blockchain, along with their corresponding public keys.\footnote{More precisely, the hash160, which is derived from the public key.} Next, we converted these hash160s into the commonly used Bitcoin address: the base-58 strings seen everywhere, on \texttt{blockchain.info}, in your wallet, etc.  We then used a shell script to manipulate the text to sort these addresses, hash160s, and their corresponding balances into a tab-delimited file. An altcoin developer could simply use this file, embed it in the first non-genesis block, then allow users to claim their coins through our coin verification process.

To create the snapshot file, we use Znort's blockchain parser\footnote{\texttt{https://github.com/znort987/blockparser}}, modified to use memory-lean hash tables.  We then use a shell script\footnote{\texttt{https://github.com/AugurProject/SideCoin/blob/master/snapshot}} to call the appropriate functions on the block parser. We specify the maximum block number as well as the number of addresses to include in the snapshot; a value of about 2,000,000 includes all addresses with balances\footnote{￼The balances are given in Satoshis.} above about $\$0.20$.  Next, the hash160s are converted into addresses\footnote{\texttt{http://bitcoin.stackexchange.com/questions/5021/how-do-you-get-the-op-hash160-value-from-a-bitcoin-address}}, which is also written to the snapshot file.  After parsing, a \texttt{snapshotToImport.txt} file is output containing the balance, hash160, and address.  An example entry is:
\begin{lstlisting}[language=json]
   180893019187.00000000 8c1d15231afa4868330f8af694ba637b69fdc2d7 14CKu2rJN2f8fdPrtmRLWChhXESgN5qaA7
\end{lstlisting}
The snapshot file has as many rows as the number of addresses we specified to the blockparser.

\section{Incorporating a Snapshot into a Blockchain's Initial State}

Once the snapshot file has been created, it must to be recorded on the new blockchain. We incorporate the snapshot data as a series of unspent transaction outputs into block one of the new blockchain. Then, we add a claim transaction function and make it available via RPC call.  This allows bitcoin owners to trustlessly claim their coins in the spinoff blockchain. Finally, we discuss how to create your own snapshot and incorporate it in a spinoff.

First, we mine the genesis block. Next, we incorporate the snapshot data into the first block.\footnote{The genesis block is loaded from memory every time \texttt{bitcoind} is started.  However, aside from the genesis block, only the last one-hundred blocks are loaded into memory. Since the snapshot block is so large, we do not want it loaded into memory every time the coin daemon is called!}  We begin by creating coinbase transactions\footnote{\texttt{http://bitcoin.org/en/developer-guide\#transaction-data}} for each of the addresses in the snapshot:
\begin{lstlisting}[language=json]
   Input:
      <pubKeyHash from snapshot>
   Output:
      Value: <Bitcoin address's balance>
      scriptPubKey: OP_DUP OP_HASH160 <pubKeyHash> OP_EQUALVERIFY OP_CHECKSIG
\end{lstlisting}
where the Bitcoin address's balance is the balance recorded in the snapshot.  After these transactions are loaded into block one, we find the block's hash, process the block (i.e., the same thing a miner does upon finding a new block), then submit it to the network.

Finally, we provide a trustless method of claiming these transactions.  Claim transactions give our users who own bitcoins a way of claiming their portion of sidecoins.  To do this, the user must submit a \emph{claim transaction}:
\begin{lstlisting}[language=json]
   Input:
      <previous transaction ID>
      <index of the output being claimed> (should be 0)
      scriptSig: <signature> <pubKey>
\end{lstlisting}
where \texttt{<pubKey>} is the public key of the user's Bitcoin address.  This input Script has successfully unlocked the new coins on the Sidecoin network!  These coins are locked up in an address belonging to the user's Sidecoin wallet:
\begin{lstlisting}[language=json]
   Output:
      Value: <Bitcoin address's balance>
      scriptPubKey: OP_DUP OP_HASH160 <Sidecoin pubKeyHash> OP_EQUALVERIFY OP_CHECKSIG
\end{lstlisting}

To facilitate submitting these claim transactions, we created a new RPC command, \texttt{claimtx}.  Users call this by inputting \texttt{claimtx} followed by the user's Bitcoin address:
\begin{lstlisting}[language=json]
   claimtx "1F1tAaz5x1HUXrCNLbtMDqcw6o5GNn4xqX"
\end{lstlisting}
The \emph{claimtx} function (in \texttt{rpcserver.cpp}) searches through the snapshot file to find the address's corresponding hash160 and unspent output in block one.  Then, it automatically builds the claim transaction (described above), locking the coins in an output belonging to a new address in the user's Sidecoin wallet:
\begin{lstlisting}[language=json]
   signrawtransaction
   '0100000001ec70650bf05a75d62f3bcf83d183064e118d39070c2a7237d33ee9a4d4930da20000000000ffffffff01606b042a010000001976a914345cd34789f945f0cbd952ce254bf5246e63be0c88ac00000000' '[{"txid":"a20d93d4a4e93ed337722a0c07398d114e0683d183cf3b2fd6755af00b6570ec","vout":0,"scriptPubKey":"76a914f4dfe70369cc98dc8113bba69bda324e4f1386a088ac"}]'
\end{lstlisting}

Next, the user is given instructions to open the RPC console of their Bitcoin client containing the address which held the bitcoins the user is claiming at the time the snapshot was taken.  The user pastes in the output supplied by the \texttt{claimtx}, which is the unsigned transaction appended to the RPC command \texttt{signrawtransaction}. The user's bitcoin client will then sign this transaction.

Finally, the user takes the results, appends them to the command \texttt{sendrawtransaction}, and submits it using Sidecoin's RPC console:
\begin{lstlisting}[language=json]
   sendrawtransaction "01000000017b1eabe0209b1fe794124575ef807057c77ada213\
                       8ae4fa8d6c4de0398a14f3f00000000494830450221008949f0\
                       cb400094ad2b5eb399d59d01c14d73d8fe6e96df1a7150deb38\
                       8ab8935022079656090d7f6bac4c9a94e0aad311a4268e082a7\
                       25f8aeae0573fb12ff866a5f01ffffffff01f0ca052a0100000\
                       01976a914cbc20a7664f2f69e5355aa427045bc15e7c6c77288\
                       ac00000000" 
\end{lstlisting}
The user has now unlocked their sidecoins! This transaction is like any other transaction: it just spends unspent outputs.  The key difference is that it involves data from two different blockchains!  We do the claim transaction in this manner to trustlessly verify that the user owns the Bitcoin private key associated with the address to which the sidecoins have been assigned.

In order to create a snapshot for use on one's own computer, one simply runs the snapshot script which asks the user a few questions about which block they'd like to take the snapshot in and how many addresses they'd like in the snapshot. After this, take the \texttt{snapshotToImport.txt} file and paste it in the \textit{Appdata/Sidecoin/balances} folder on your respective OS. Place a \texttt{sidecoin.conf} file there, as well. Open the \texttt{snapshot.h} file in Sidecoin's source folder and set \texttt{GENESIS\_SWITCH} to true. Then, launch Sidecoin, and it will mine the mainnet, testnet, and regtest genesis blocks.  (Upon completion, the block hashes and Merkle root fields in \texttt{chainparams.cpp} need to be updated.)

After the genesis block is mined, launch Sidecoin again and it will load the transactions into block one from the snapshot file, then it will add block one to the blockchain.  If it outputs \texttt{accepted} everything has been done correctly! Next, change \texttt{GENESIS\_SWITCH} to false.  Finally, relaunch Sidecoin to see a new successful spinoff implementation. 

For users who want to use an already created spinoff coin, one can simply wait to download block one from other peers. However, when the network is first starting off this is very slow and impractical due to the large size of block one. Instead, one can try to load block one in on their own by enabling the \texttt{GENESIS\_SWITCH} and editing the \texttt{blockOne.nNonce} and \texttt{nTime} to the values that they were mined with if not already set. Then place the \texttt{snapshotToImport.txt} file in the \emph{balances} directory, start Sidecoin, wait for it to load the block, then disable the \texttt{GENESIS\_SWITCH}.\footnote{The text file is needed for anyone looking to claim a transaction anyway, so this a nice method.  An alternative method is to paste the blocks and chainstate folders from a peer who has a blockchain with block one in it into one's Sidecoin app data folder. Block one is stored as a checkpoint, so there is not much concern of someone falsifying the snapshot and giving a faulty block one.}

\section{Proof-of-Concept}

Sidecoin\footnote{\texttt{https://github.com/AugurProject/SideCoin}} is our proof-of-concept implementation of a snapshot/spinoff mechanism.  It is a fork of Bitcoin Core 0.9.1.  It does not support pay-to-script-hash addresses in the snapshot, nor does it support simplified claim verification schemes, as proposed on Bitcointalk.\footnote{\texttt{http://bitcointalk.org/index.php?topic=563972.0}}

We made a few changes to our fork of Bitcoin Core to make Sidecoin. We had to change a couple things in the \texttt{CheckBlock} code segments: allowing multiple coinbase transactions, as well as modifying block size limits, signature operations limits, and transaction limits. We only made these modifications for block one, which is determined as the block with a previous block hash equivalent to the genesis block.  We also added functions allowing Bitcoin to read in snapshot data from a flat file into block one, the \texttt{claimtx} function, and manually loading block one into the blockchain, similar to how the genesis block is built.￼

Finally, we recognize that most of the utility of snapshots and spinoffs will ultimately be replaced by the interoperability provided by Sidechains.\footnote{\texttt{http://www.blockstream.com/sidechains.pdf}}  However, for those looking to make an altcoin and dramatically boost community adoption, spinoffs could be a solution.  Another possible application of the snapshot is using it to bootstrap a replacement for Bitcoin in the unlikely event something catastrophic occurs.

\acknowledgements{Financial support for this project was provided by Joe Costello.}

\end{document}